\documentclass{article}
\usepackage{times}
\usepackage{amsfonts}
\usepackage{amssymb}
\usepackage{graphicx}
\begin{document}
\noindent
{\Large EMERGENT QUANTUM MECHANICS AS A THERMAL ENSEMBLE}\\
\vskip1cm
\noindent
{\bf  P. Fern\'andez de C\'ordoba$^{1,a}$, J.M. Isidro$^{1,b}$ and Milton H. Perea$^{1,2,c}$}\\
${}^{1}$Instituto Universitario de Matem\'atica Pura y Aplicada,\\ Universidad Polit\'ecnica de Valencia, Valencia 46022, Spain\\
${}^{2}$Departamento de Matem\'aticas y F\'{\i}sica, Universidad Tecnol\'ogica\\ del Choc\'o, Colombia\\
${}^{a}${\tt pfernandez@mat.upv.es}, ${}^{b}${\tt joissan@mat.upv.es}\\
${}^{c}${\tt milpecr@posgrado.upv.es}  \\
\vskip.5cm
\noindent
%\today
%\vskip.5cm
\noindent
{\bf Abstract} It has been argued that gravity acts dissipatively on quantum--mechanical systems, inducing thermal fluctuations that become indistinguishable from quantum fluctuations. This has led some authors to demand that some form of time irreversibility be incorporated into the formalism of quantum mechanics. As a tool towards this goal we propose a thermodynamical approach to quantum mechanics, based on Onsager's classical theory of irreversible processes and on Prigogine's nonunitary transformation theory. An entropy operator replaces the Hamiltonian as the generator of evolution. The canonically conjugate variable corresponding to the entropy is a dimensionless evolution parameter. Contrary to the Hamiltonian, the entropy operator is not a conserved Noether charge. Our construction succeeds in implementing gravitationally--induced irreversibility in the quantum theory.

%\tableofcontents
\section{Introduction}\label{uno}

It has been known for long that weak interactions violate CP--invariance \cite{CRONIN}. By the CPT theorem of quantum field theory, time invariance must also be violated in weak interactions; recent observations \cite{BABAR} confirm this expectation. Now quantum field theory is an extension of quantum mechanics. Since time invariance is naturally implemented in the latter, it would appear that only CP--violating quantum field theories can also violate time invariance, because quantum mechanics as we know it is symmetric under time reversal.

Actually such is not the case. A number of firmly established quantum--gravity effects have been shown to be intrinsically irreversible;  for background see,  {\it e.g.}\/,  \cite{THOOFT1, KIEFER, WALD1, WALD2, ZEH} and references therein. From the independent perspective of statistical physics \cite{PRIGOGINE} it has also been suggested that time irreversibility should be taken into account at the more fundamental level of the differential equations governing mechanical processes. This is in sharp contrast with standard thinking, where irreversibility is thought to arise through {\it time--irreversible}\/ initial conditions imposed on the solutions to {\it time--reversible}\/ evolution equations. In view of this situation, a number of authors have called for the due modifications to the standard quantum--mechanical formalism (for a detailed account and original references see, {\it e.g.}\/, \cite{PENROSE2}). Specifically, in this paper we tackle the problem of incorporating some form of time irreversibility at the level of the differential equation governing evolution  \cite{PRIGOGINE}.

Closely related to this viewpoint is the {\it emergent approach}\/ to physics. The latter has been the subject of a vast literature (see  \cite{CARROLL} for a comprehensive review),  but let us briefly mention some noteworthy aspects. The notion of an {\it emergent theory}\/, that is, the concept that a given physical theory could be an effective model of some deeper--level degrees of freedom, has been postulated of a number of existing theories, most notably of gravity and of quantum mechanics.  In the particular case of the latter, refs. \cite{ADLER1, ELZE1, ELZE2, HOLLOWOOD, THOOFT2, THOOFT3, THOOFT5, SMOLIN3} address this issue from a number of different perspectives. The paradigm that {\it quantisation is dissipation}\/, implicitly present in some of the above approaches, has been made precise in \cite{BLASONE00,  BLASONE2}\/.  Frequently, these takes on quantum physics can be completely recast in purely classical terms \cite{MATONE1, KHRENNIKOV, WETTERICH}.  An alternative perspective, based on classical nonequilibrium thermodynamics \cite{ONSAGER}, has been advocated in \cite{NOI1, NOI3, NOI4}. Beyond quantum mechanics, the relevance of nonequilibrium physics for quantum gravity and strings has been emphasised recently \cite{MINIC, HU1}.

The basic physical assumption we will make use of posits that {\it spacetime is not a fundamental concept, but rather an emergent phenomenon instead}\/. In fact this hypothesis is not at all new (for references and background see, {\it e.g.}\/, \cite{KIEFER}), some of its most recent incarnations being \cite{PADDY1, PADDY2, VERLINDE}. Once spacetime is no longer regarded as a fundamental concept, but rather as a derived notion, then every theory that makes use of spacetime concepts automatically qualifies as emergent. Such is the case of quantum mechanics. For our purposes it will suffice to concentrate on the time variable and expose its emergent nature. We will therefore try to express time in terms of thermodynamical quantities, and explore the consequences for the quantum theory. Again, the notion of time as having a thermodynamical origin is not new \cite{DEBROGLIE, CONNES}, having reappeared more recently in \cite{ROVELLI3, ROVELLI0, ROVELLI1, ROVELLI2}; see also \cite{ELLIS, MONTEVIDEO, HARTLE, HILEY} for related views. New to our approach is the notion that {\it an emergent time variable automatically implies that quantum theory itself qualifies as an emergent phenomenon}\/. Specifically, the possibility of reexpressing the nonrelativistic Schroedinger equation in purely entropic terms (instead of its usual Hamiltonian language) implies that quantum mechanics involves some degree of coarse graining of microscopic information. In our approach, the very existence of an entropy operator replacing the Hamiltonian operator is an inequivocal clue of this coarse graining. 

To begin with, we would like to draw attention to the following analogy. On one hand we have the quantum--mechanical time--energy uncertainty relation
\begin{equation}
\Delta E\Delta t\gtrsim \hbar.
\label{incertezza}
\end{equation}
On the other hand, in the theory of irreversible thermodynamics \cite{ONSAGERRECIP, ONSAGER}, one computes the average product of the fluctuations of the entropy and the temperature for a thermodynamical system slightly away from equilibrium (this is the linear regime, also called the Gaussian approximation). This product turns out to be given by \cite{LANDAU}
\begin{equation}
\Delta S\Delta T = k_BT,
\label{termolandau}
\end{equation}
$k_B$ being Boltzmann's constant. The change of variables 
\begin{equation}
\tau:=\ln \left(\frac{T}{T_0}\right), 
\label{definito}
\end{equation}
where $T_0$ is some reference temperature, reduces (\ref{termolandau}) to 
\begin{equation}
\Delta S\Delta\tau \gtrsim k_B.
\label{landautermos}
\end{equation}
In (\ref{landautermos}) we have taken the liberty of replacing the equality sign of (\ref{termolandau}) with an inequality; the latter is saturated in the Gaussian approximation (used in the derivation of (\ref{termolandau})). Beyond the Gaussian regime, one expects the inequality to hold strictly. As we will see, the analogy between (\ref{incertezza}) and (\ref{landautermos}) is more than just a happy coincidence---it is in fact anything but accidental.

\section{Emergent time}\label{dos}

Let $t$ and $T$ respectively denote nonrelativistic time and absolute temperature, as measured by an inertial observer that will be kept the same throughout. We posit that $t^{-1}$ equals $T$ modulo dimensional factors:
\begin{equation}
\frac{C}{t}=\frac{k_B}{\hbar}T.
\label{tiemponuevo}
\end{equation}
Here $C$ is a dimensionless numerical factor, whose value we will pick presently in order to suit our needs. Modulo this $C$, which will play a prominent role in what follows, the relation  (\ref{tiemponuevo}) between time and temperature was postulated long ago by de Broglie \cite{DEBROGLIE}. A related change of variables has been used more recently in \cite{RUUGE}.

Beyond purely dimensional grounds, there are deeper motivations for Eq. (\ref{tiemponuevo}). Specifically, in \cite{NOI3, NOI4} we have established a map between  quantum mechanics (in the Gaussian approximation) and the classical theory of irreversible thermodynamics (in the linear regime).\footnote{As argued in \cite{NOI3, NOI4}, the linear regime in irreversible thermodynamics is the analogue of the semiclassical, or Gaussian, approximation to quantum mechanics.} In this latter theory \cite{ONSAGER} we have $N$ independent thermodynamical coordinates $y^1, \ldots, y^N$ on which the entropy $S$ depends, and $N$ conjugate forces $Y_k:={\partial S}/{\partial y^k}$. Let $t'$ denote thermodynamical time. The assumption of linearity between the velocities $\dot y^k$ and the forces $Y_j$ amounts to
\begin{equation}
\dot y^i=\frac{{\rm d}y^i}{{\rm d}t'}=\sum_{j=1}^NL^{ij}Y_j,\qquad  Y_i=\sum_{j=1}^NR_{ij}\dot y^j, \qquad R_{ij}=(L^{ij})^{-1}.
\label{flajos}
\end{equation}
Under the assumption that the underlying microscopic dynamics is time--reversible, the constant matrix $L_{ij}$ turns out to be symmetric (Onsager's reciprocity theorem) \cite{ONSAGERRECIP}. By (\ref{flajos}), the time rate of entropy production can be written either as a quadratic form in the velocities, or as a quadratic form in the forces:
\begin{equation}
\dot S=\sum_{i,j=1}^NR_{ij}\dot y^i\dot y^j=\sum_{i,j=1}^NL^{ij}Y_iY_j.
\label{taxa}
\end{equation}
We see that it is not the entropy $S$, but its time rate of production $\dot S$, that plays the role of a (harmonic) Hamiltonian, because\footnote{$L^{ij}$ is positive definite for a dissipative process, hence also $R_{ij}$.}
\begin{equation}
\dot S=\frac{{\rm d}S}{{\rm d}t'}=\frac{1}{2}\sum_{i,j=1}^N\left(R_{ij}\dot y^i\dot y^j+L^{ij}Y_iY_j\right).
\label{porque}
\end{equation}
Here again we see that inverse time can be regarded as temperature. In Eqs. (\ref{flajos})--(\ref{porque}) above, the thermodynamical time $t'$ and the mechanical time $t$ are related as per the Wick rotation, $t'={\rm i}t$  \cite{NOI3, NOI4}.  Thus we expect a thermodynamical approach to quantum mechanics to involve the complexification of time. Multiplying (\ref{tiemponuevo}) through by $H/T$, one realises that (\ref{tiemponuevo}) is roughly equivalent to
\begin{equation}
C\frac{{\rm d}S}{{\rm d}t}=\frac{k_B}{\hbar}H,
\label{bruecke}
\end{equation}
which bridges the gap between the mechanical point of view (the right--hand side of (\ref{bruecke})) and the thermodynamical point of view (the left--hand side).  The above is a handwaving argument to justify equating the time variation of the entropy with the energy (modulo dimensional constants); we will actually {\it derive}\/ Eq. (\ref{bruecke}) later on (see (\ref{noconservada})).  Eq. (\ref{bruecke}) is also important because it holds beyond its Gaussian limit given in (\ref{porque}). In what follows we will work out in detail the relationship between the mechanical and the thermodynamical points of view expressed above.

\section{Entropy {\it vs}\/. energy}\label{tres}

\subsection{The energy picture}\label{papst}

{}For reasons that will become apparent presently let us call quantum mechanics, in its standard formulation, the {\it energy picture}\/ of quantum mechanics; we will also use the term {\it $H$--picture}\/.\footnote{We use the term {\it picture}\/ instead of its synonym {\it representation}\/ in order to avoid confusion with the technical meaning of the latter term in quantum--mechanical contexts such as {\it choice of basis in Hilbert space}\/, or {\it group representation}\/, or similar. Expressions such as {\it Schroedinger picture}\/, or {\it Heisenberg picture}\/, or related terms used in standard quantum mechanics should also not be confused with our use of the word {\it picture}\/.} The evolution of pure quantum states is governed by the Schroedinger equation,
\begin{equation}
{\rm i}\hbar\frac{{\rm d}\psi}{{\rm d}t}=H\psi.
\label{chro}
\end{equation}
The general solution to the above reads $\psi(t)={\cal U}(t)\psi(0)$, where
\begin{equation}
{\cal U}(t):={\cal T}\exp\left(-\frac{{\rm i}}{\hbar}\int_0^tH(\tilde t){\rm d}\tilde t\right),
\label{elephat}
\end{equation}
and ${\cal T}$ denotes the ordering operation along the evolution parameter $\tilde t$. When $t\in\mathbb{R}$, the time--evolution operators ${\cal U}(t)$ define a 1--parameter group of unitary operators that ensure the reversibility of time flow in the $H$--picture.

\subsection{The entropy picture}\label{babst}

The purpose of this section is to develop {\it the entropy picture}\/ of quantum mechanics, or the {\it $S$--picture}\/ for short.

Under the combined changes of variables (\ref{tiemponuevo}) and (\ref{definito}), the evolution equation (\ref{chro}) becomes
\begin{equation}
-\frac{{\rm i}k_B}{C}\frac{{\rm d}\psi}{{\rm d}\tau}=S\psi,
\label{entrochro}
\end{equation}
where we have defined the entropy operator $S$ 
\begin{equation}
S:=\frac{H}{T}.
\label{chrotis}
\end{equation}
The new evolution parameter $\tau$ is dimensionless, while $S$ carries the dimension of an entropy. Our time variable $\tau$ coincides with the thermal time of \cite{CONNES, ROVELLI0, ROVELLI1}, the latter specified to the nonrelativistic limit correponding to the Schroedinger wave equation. We will see presently that $C\in\mathbb{C}$, so our evolution variable $\tau$ will actually be a complexified (or Wick--rotated), nonrelativistic, dimensionless, thermal--time variable.

The solution to the evolution equation (\ref{entrochro}) can be written as
\begin{equation}
\psi(\tau)={\cal S}_{C}(\tau)\psi(0),\qquad \tau\geq 0,
\label{mala}
\end{equation} 
where
\begin{equation}
{\cal S}_{C}(\tau):={\cal T}\exp\left(\frac{{\rm i}C}{k_B}\int_0^{\tau}S(\tilde\tau){\rm d}\tilde\tau\right)
\label{ebolution}
\end{equation}
and ${\cal T}$ denotes the ordering operation along the the evolution parameter $\tilde\tau$. If we now pick $C\in\mathbb{R}$, the evolution operators $\{{\cal S}_{C}(\tau),\tau\in\mathbb{R}\}$ in (\ref{ebolution}) form a 1--parameter group of unitary operators. 

As long as $C$ remains real, Eqs. (\ref{entrochro})--(\ref{ebolution}) above simply restate standard quantum mechanics using the alternative set of variables $(\tau, S)$. It is only for $C\notin\mathbb{R}$ that time evolution can become irreversible. For this purpose let us set, dropping an irrelevant real normalisation,
\begin{equation}
C:={\rm e}^{{\rm i}\varphi},\qquad \varphi\in\mathbb{R}.
\label{choix}
\end{equation}
On the complex plane, (\ref{choix}) corresponds to Wick--rotating the time axis by an angle $\varphi$. Now certain special values of $\varphi$ are known to correspond to specific physical situations. For example, $\varphi=0$ corresponds to standard quantum mechanics, while $\varphi=\pi$ implements the time reverse of $\varphi=0$. The value $\varphi=-\pi/2$ gives a positive real argument within the exponential of (\ref{ebolution}); we will see in section \ref{hacheese} that this corresponds to the case of maximal entropy production, or maximal dissipation. Finally, the value $\varphi=\pi/2$ gives a negative real argument within the exponential of (\ref{ebolution}); this will turn out to correspond to the unphysical situation of maximal {\it anti}\/dissipation. All other values of  $\varphi$ therefore correspond to intermediate situations between exactly unitary evolution (eventually, time--reversed) and maximal dissipation (eventually, antidissipation). For obvious reasons we must pick the quadrant corresponding to the forward time direction and positive dissipation, {\it i.e.}\/, $\varphi\in[-\pi/2, 0]$. Let the dimensionless variable $x\in\mathbb{R}$ be a measure of the external gravitational field acting on the particle of mass $m$ described by the Hamiltonian $H$, such that $x=0$ describes the absence of gravitation, and $x\to\infty$ describes the case of a strong gravitational field acting on $m$. From what is known concerning the effects of gravitational fields on the quantum mechanics of particles we expect the phase $\varphi$ to depend on $x$ roughly as follows:
\begin{equation}
\varphi(x)=-\frac{\pi}{2}\left(1-{\rm e}^{-x}\right), \qquad x\geq 0.
\label{moddell}
\end{equation}
Indeed, for $x=0$ we have a perfectly unitary evolution ($\varphi=0$) as befits quantum particles in the absence of gravitation, while for strong gravitational fields ($x\to\infty$) we have $\varphi\to-\pi/2$, and unitarity gives way to dissipation. Of course, the precise profile (\ref{moddell}) for the function $\varphi(x)$ is just one out of many possible, but it captures the right physical behaviour, namely,  that gravitational fields induce thermal dissipative effects in the quantum theory, in such a way as to render quantum uncertainties indistinguishable from statistical fluctuations \cite{SMOLIN1, SMOLIN2}. In the absence of a gravitational field, any inertial observer perceives a clear--cut separation between these two types of fluctuations.

Altogether, (\ref{choix}) and (\ref{moddell}) yield
\begin{equation}
C(x)=\exp\left[-\frac{{\rm i}\pi}{2}\left(1-{\rm e}^{-x}\right)\right].
\label{choixx}
\end{equation}
{}For the rest of this paper we will concentrate on the limiting case of a weak gravitational field. So we have\footnote{We will henceforth drop terms of order $\varepsilon^2$ and higher.}
\begin{equation}
C(\varepsilon)\simeq 1+{\rm i}\varepsilon,\qquad \varepsilon=-\frac{\pi x}{2}, \qquad x\geq 0. 
\label{chois}
\end{equation}
It remains to identify a dimensionless variable $x$ that can provide a physically reasonable measure of a weak gravitational field acting on the quantum particle.\footnote{In a sense, the situation analysed here is complementary to that described in ref. \cite{KENT}.} It is standard to parametrise such a field by the metric $g_{\mu\nu}=\eta_{\mu\nu}+h_{\mu\nu}$, where $\eta_{\mu\nu}$ is the Minkowski metric, and $h_{\mu\nu}$ a small correction. It is also convenient to introduce the quantities $h^{\lambda}_{\mu}:=\eta^{\lambda\alpha}h_{\mu\alpha}$ and $h:=h^{\alpha}_{\alpha}=\eta^{\sigma\lambda}h_{\sigma\lambda}$. The linearised Einstein equations read 
\begin{equation}
-16\pi T^{\nu}_{\mu}=\eta^{\sigma\lambda}\frac{\partial^2}{\partial x^{\sigma}\partial x^{\lambda}}\left(h^{\nu}_{\mu}-\frac{1}{2}\eta^{\nu}_{\mu}h\right),
\label{conton}
\end{equation}
and we can take $x=\langle h\rangle$ as a variable that satisfies our needs, at least in the weak field limit considered here. The angular brackets in $\langle h\rangle$ stand for the average value of the function $h$ over the spacetime region of interest. That $\langle h\rangle$ is nonnegative follows from the fact that \cite{TOLMAN}
\begin{equation}
h=4\int\frac{[T^\alpha_{\alpha}]}{r}{\rm d}x{\rm d}y{\rm d}z,\qquad T^\alpha_{\alpha}\geq 0.
\label{tarde}
\end{equation}
The square brackets around the trace $T^\alpha_{\alpha}$ stand for the evaluation at a time earlier than that of interest by the interval needed for a signal to pass with unit velocity from the element ${\rm d}x{\rm d}y{\rm d}z$ to a point a distance $r$ apart.

Substitution of (\ref{chois}) into (\ref{ebolution}) leads to
\begin{equation}
{\cal S}_{1+{\rm i}\varepsilon}(\tau):={\cal T}\exp\left(\frac{{\rm i}-\varepsilon}{k_B}\int_0^{\tau}S(\tilde\tau){\rm d}\tilde\tau\right),
\label{iebolution}
\end{equation}
and the set $\left\{{\cal S}_{1+{\rm i}\varepsilon}(\tau), \tau\geq 0\right\}$ forms a 1--parameter {\it semigroup}\/ of nonunitary operators. 
In the limit $\varepsilon=0$, the set $\left\{{\cal S}_{1}(\tau), \tau\in\mathbb{R}\right\}$ becomes again the 1--parameter group of unitary operators given in $(\ref{ebolution})$ (with $C=1$). The parameter $\varepsilon$ allows for a continuous transition between the unitary ($\varepsilon=0$)  and the nonunitary ($\varepsilon\neq 0$) regimes. 

Our choice (\ref{chois}) yields in (\ref{entrochro})
\begin{equation}
 -({\rm i}+\varepsilon)k_B\frac{{\rm d}\psi}{{\rm d}\tau}= S\psi.
\label{maniare}
\end{equation}
It makes sense to call (\ref{maniare}) the {\it entropic Schroedinger equation}\/. Again, in the limit $\varepsilon=0$ we recover a Schroedinger--like equation,
\begin{equation}
-{\rm i}k_B\frac{{\rm d}\psi}{{\rm d}\tau}=S\psi.
\label{mangiarebene}
\end{equation}
The $\varepsilon$ term on the left--hand side of (\ref{maniare}) can be regarded as a perturbative correction to the derivative term in (\ref{mangiarebene}). We see that it breaks unitarity explicitly, already at the level of the differential equation governing evolution. The physical reason for this breakdown of unitarity is the presence of an external gravitational field, the strength of which is parametrised by $\varepsilon$.

Altogether, Eqs. (\ref{iebolution}) and (\ref{maniare})  define the $S$--picture of quantum mechanics.

\subsection{$S$ rather than $H$}\label{hacheese}

One might argue that there is no need for the $S$--picture because the $H$--picture suffices. Indeed it has been known for long that a simple, ``phenomenological" implementation of nonunitarity within the $H$--picture consists in the addition of a nonvanishing imaginary part to the time variable $t$ in (\ref{chro}): 
\begin{equation}
({\rm i}+\varepsilon')\hbar\frac{{\rm d}\psi}{{\rm d}t}=H\psi.
\label{chrodelta}
\end{equation}
Here $\varepsilon'\in\mathbb{R}$ is a small (dimensionless) perturbation.  What distinguishes (\ref{chrodelta}) from its entropic partner (\ref{maniare}), and why is the latter to be preferred over the former? 

In terms of the variables $(t, H)$, invariance under translations in $t$ is reflected in the conservation of the Noether charge $H$. There exists no  preferred origin $t=0$ for time. While (\ref{chrodelta}) certainly leads to energy dissipation, the natural physical quantity to describe dissipation is the entropy, where one expects to find ${\rm d}S/{\rm d}t\geq 0$ instead of a conservation law. In the variables $(\tau, S)$ of (\ref{maniare}), one expects to have no conservation law at all; one actually finds\footnote{Here we are assuming ${\rm d}H/{\rm d}t= 0$ for simplicity.}
\begin{equation}
\frac{{\rm d}S}{{\rm d}t}=\frac{k_B}{\hbar}(1-{\rm i}\varepsilon)H,
\label{noconservada}
\end{equation}
as anticipated in (\ref{bruecke}). Now, from (\ref{porque}) and the Wick rotation $t'={\rm i}t$, we conclude that it is ${\rm Im}\left({\rm d}S/{\rm d}t\right)$, and not ${\rm Re}\left({\rm d}S/{\rm d}t\right)$, that accounts for dissipation. Indeed, recalling (\ref{tiemponuevo}), the real part of  (\ref{noconservada}) is the usual thermodynamical definition of temperature, $\partial S/\partial E=1/T$. In other words, even if ${\rm Re}\left({\rm d}S/{\rm d}t\right)=k_BH/\hbar \neq 0$, this latter equation alone does not account for dissipation. Since 
\begin{equation}
{\rm Im}\left(\frac{{\rm d}S}{{\rm d}t}\right)=-\varepsilon\frac{k_B}{\hbar}H,
\label{disipatiu}
\end{equation}
there will be no conservation law for $S$ under evolution in $t$ if $\varepsilon\neq 0$. The same conclusion applies to evolution in $\tau$. Furthermore, dissipation vanishes in the limit $\varepsilon=0$ as had to be the case. Finally, for Eq. (\ref{disipatiu}) to be consistent with the second law of thermodynamics, we need to choose $\varepsilon<0$, as anticipated in (\ref{chois}). This latter point is obvious in the Gaussian approximation (\ref{porque}), where $H$ is a positive--definite quadratic form, but it also holds true beyond that approximation, because $H$ is bounded from below (if needed, one adds a constant to shift the energy of the groundstate, to make it nonnegative).

As already remarked, the operators (\ref{iebolution}) are unitary iff $\varepsilon=0$. Here we see that their nonunitarity differs considerably in the two cases $\varepsilon>0$ and $\varepsilon<0$. Since $\tau\geq 0$, had $\varepsilon$ been positive, this would have turned the ${\cal S}_{1+{\rm i}\varepsilon}(\tau)$ into a semigroup of {\it contraction operators}\/ \cite{YOSIDA}, which would describe an unphysical {\it anti}\/dissipative world. On the contrary, the choice $\varepsilon<0$ of (\ref{chois}) leads to the opposite behaviour,  {\it dilatation}\/, which is in agreement with the second law of thermodynamics. 

In the $H$--picture, whenever the Hamiltonian is time--independent, there exist energy eigenstates $\phi$ satisfying $H\phi=E\phi$; the wavefunction $\psi$ then factorises as $\psi=\phi\exp(-{\rm i}Et/\hbar)$. A similar property holds in the $S$--picture, assuming that $H$ remains $t$--independent, hence also $\tau$--independent. In this latter case one can readily check that the factorised wavefunctions
\begin{equation}
\psi=\phi\,{\rm e}^{({\rm i}-\varepsilon)\tau s},
\label{integro}
\end{equation}
where $\phi$ does not depend on $\tau$, lead to the eigenvalue equation
\begin{equation}
S\phi=sk_B\phi,
\label{autoff}
\end{equation}
with $s\in\mathbb{R}$ playing the role of a dimensionless entropic eigenvalue. Again, eqs. (\ref{integro}) and (\ref{autoff}) above are in perfect agreement with the second law of thermodynamics.

To summarise, unitarity is violated in the $S$--picture, where $\varepsilon <0$ appears, but not in the $H$--picture, where the evolution equations (\ref{chro}) and (\ref{elephat}) remain strictly valid. As such, this ``change of picture" between $H$ and $S$ is an instance of Prigogine's {\it nonunitary transformation}\/ \cite{PRIGOGINE}. The apparent dilemma, ``Is unitarity violated or not?", will be resolved in section \ref{riemsphti}.

\subsection{Uncertainty {\it vs}\/. the second law}\label{cuatro}

It is common lore that, at least for large enough temperatures, quantum fluctuations are negligible compared to thermal statistical fluctuations \cite{LANDAU}. 
When stating that, {\it in the presence of a gravitational field}\/, quantum fluctuations are inextricably linked with thermal statistical fluctuations, one is postulating a new kind of uncertainty principle: {\it the indistinguishability between quantum and statistical fluctuations}\/ \cite{CATICHA, SMOLIN1, SMOLIN2}. Here we will provide an example of this indistinguishability. A look at Eq. (\ref{incertezza}) and a comparison of (\ref{maniare}) with (\ref{chro}) leads one to conclude the following uncertainty relation:
\begin{equation}
\Delta S\Delta \tau\gtrsim k_B.
\label{incierto}
\end{equation}
It is rewarding to see the product of {\it thermal fluctuations}\/ found in (\ref{landautermos}) nicely matched by the product of {\it quantum--mechanical uncertainties}\/ (\ref{incierto}). This is more than just a coincidence---it is an expression of the fact that, {\it in the presence of a gravitational field}\/, quantum uncertainties can be understood as statistical fluctuations possessing a thermal origin \cite{SMOLIN1, SMOLIN2}. The above uncertainty relation leads to the factor $2k_B$ replacing the quantum of action $\hbar$, in perfect agreement with the results of \cite{RUUGE}.

Since $\tau$ is dimensionless, we can safely set $\Delta\tau=1$ in (\ref{incierto}) with the certainty that this numerical value will not change upon changing units. This leads to
\begin{equation}
\Delta S\geq k_B> 0,
\label{segunda}
\end{equation}
which becomes the familiar second law of thermodynamics when written as
\begin{equation}
\Delta S\geq 0.
\label{lex}
\end{equation}
Strictly speaking, the equality in (\ref{lex}) is never attained, as $k_B>0$. However, in the limit $k_B\to 0$ we can saturate the inequality in (\ref{lex}) and have $\Delta S=0$. The limit $k_B\to 0$ has been argued to correspond to the semiclassical limit $\hbar\to 0$ of quantum mechanics \cite{NOI1}.\footnote{In order to conform to the conventions of ref. \cite{VERLINDE}, in ref. \cite{NOI1} we have normalised the quantum of entropy to the value $2\pi k_B$ instead of the value $2k_B$ used here.}

We conclude that the quantum--mechanical uncertainty principle provides the refinement (\ref{segunda}) of the second law of thermodynamics (\ref{lex}), to which it becomes strictly  equivalent in the semiclassical limit $k_B\to 0$.

\subsection{Commutators {\it vs}\/. fluctuations}\label{comufluk}

In the standard quantum--mechanical formalism, nonvanishing commutators account for uncertainties. Fortunately for us, uncertainties can arise from fluctuations  just as well as from commutators. In keeping with our previous arguments, here we will take statistical fluctuations as our starting point, in order to arrive at commutators. 

We will illustrate our point by means of an example. Consider a thermodynamical system described by the temperature $T$, the pressure $p$, the volume $V$ and the entropy $S$. Now, in the Gaussian approximation,  the probability $P$ of a fluctuation $\Delta p$, $\Delta V$, $\Delta T$, $\Delta S$ is given by \cite{LANDAU}
\begin{equation}
P=Z^{-1}\exp\left[-\frac{1}{2k_BT}(-\Delta p\Delta V+\Delta T\Delta S)\right].
\label{probaflu}
\end{equation}
If we have an equation of state $F(p,V,T)=0$ we can solve for the temperature to obtain $T=g(p,V)$. This allows us to rewrite (\ref{probaflu}) as
\begin{equation}
P=Z^{-1}\exp\left[-\frac{1}{2k_B}\left(-\frac{\Delta p\Delta V}{g(p, V)}+\frac{\Delta T\Delta S}{T}\right)\right].
\label{fluprobaflu}
\end{equation}
This somewhat clumsy expression can be further simplified  if we assume our system to be an ideal gas, $pV=S_0T$:\footnote{Here $S_0$ is the mole number $n$ times the gas constant $R$. Whether or not our system is an ideal gas is immaterial, as the change of variables (\ref{tiempo}) can be modified appropriately without altering our conclusions.}
\begin{equation}
P=Z^{-1}\exp\left[-\frac{1}{2k_B}\left(-S_0\frac{\Delta p\Delta V}{pV}+\frac{\Delta T\Delta S}{T}\right)\right].
\label{fluflu}
\end{equation}
{}Finally define the dimensionless variables
\begin{equation}
p_1:=-\ln \left(\frac{p}{p_0}\right),\qquad q_1:=\ln \left(\frac{V}{V_0}\right),\quad p_2:=\ln \left(\frac{T}{T_0}\right), \quad q_2:=\frac{S}{S_0},
\label{tiempo}
\end{equation}
where $p_0$, $V_0$, $T_0$, $S_0$ are fixed reference values, to arrive at
\begin{equation}
P=Z^{-1}\exp\left[-\frac{S_0}{2k_B}\left(\Delta p_1\Delta q_1+\Delta p_2\Delta q_2\right)\right].
\label{muchafluflu}
\end{equation}
The argument of the above exponential is very suggestive. Indeed, let $q_1,q_2$ be coordinates on the thermodynamical configuration space $Y$, and consider the (dimensionless) symplectic form on the cotangent bundle $T^*Y$ given by
\begin{equation}
\Omega={\rm d}p_1\wedge{\rm d}q_1+{\rm d}p_2\wedge{\rm d}q_2.
\label{sindim}
\end{equation}
We have
\begin{equation}
\Omega={\rm d}\theta, \quad \theta:=p_1\,{\rm d}q_1+p_2\,{\rm d}q_2.
\label{ditat}
\end{equation}
Now $\Delta p_1\Delta q_1+\Delta p_2\Delta q_2$ equals the (symplectic) area of a 2--dimensional open surface $D$ within $T^*Y$,
\begin{equation}
\Delta p_1\Delta q_1+\Delta p_2\Delta q_2=\int_D\left({\rm d}p_1\wedge{\rm d}q_1+{\rm d}p_2\wedge{\rm d}q_2\right)=\int_D{\rm d}\theta,
\label{super}
\end{equation}
the boundary of which is $\partial D\neq 0$ (the surface $D$ can be taken to be open precisely because $D$ is caused by a fluctuation). Applying Stokes' theorem we can thus write for the probability (\ref{muchafluflu})
\begin{equation}
P=Z^{-1}\exp\left(-\frac{S_0}{2k_B}\int_D\Omega\right)
\label{chuli}
\end{equation}
$$
=Z^{-1}\exp\left(-\frac{S_0}{2k_B}\int_D{\rm d}\theta\right)=Z^{-1}\exp\left(-\frac{S_0}{2k_B}\int_{\partial D}\theta\right).
$$
Starting from fluctuations, which render commutators unnecessary in the thermodynamical description, we have arrived back at a mechanical description in terms of a symplectic form. The inverse of the latter gives Poisson brackets and, upon quantisation, commutators. This simple example illustrates the thermodynamical analogue of quantum commutators.

\subsection{Quantumness {\it vs}\/. dissipation}\label{riemsphti}

To round up our presentation of quantum theory in thermodynamical terms, let us see how suggestive Eq. (\ref{tiemponuevo}) is of a closely related geometric construction. 

Assume being given two copies of the complex plane $\mathbb{C}$, one parametrised by the complex coordinate $z$, the other by $\omega$. Then the set formed by the two coordinate charts $\left\{z\in\mathbb{C}\right\}$ and $\left\{w\in\mathbb{C}\right\}$ defines an (analytic) atlas covering the Riemann sphere $S^2$, where $z=0$ (respectively, $w=0$) corresponds to the north pole (respectively, south pole). The transition between these coordinates is $w=-1/z$, which coincides with (\ref{tiemponuevo}) up to dimensional constants.

In this way it is very tempting to identify $(t, T)$ with $(z, w)$; of course, the latter are real 2--dimensional variables, while the former are real 1--dimensional. We may thus  regard the pair ``time, temperature" as coordinates on a copy of the circle $S^1$ that one might call {\it the circle of time}\/, or {\it the circle of temperature}\/  just as well \cite{DOLCE}. Since the circle $S^1$ is a compact manifold, charting it smoothly requires at least two coordinate charts (in our case $T$ and $t$). In physical terms, temperature is the physical variable that compactifies time, and viceversa \cite{MATONE2}. The rotation (by $2\pi$ radians) of any circle $S^1$ joining the north and south poles spans the whole sphere $S^2$. This same geometrical rotation (now by an angle $\varepsilon$) corresponds to the Wick rotation of (\ref{chois}). Thus Wick--rotating the circle of time $S^1$ by all possible angles generates the whole sphere $S^2$.

Now, the $H$--picture discussed in section \ref{papst} corresponds to viewing quantum mechanics {\it in the absence of dissipation}\/. As already observed, this situation corresponds to {\it the absence of a gravitational field}\/. On the Riemann sphere $S^2$, the $H$--picture describes quantum mechanics with respect to an evolution parameter $t$ that runs over the real axis ${\rm Im}(z)=0$ within the coordinate chart $\left\{z\in\mathbb{C}\right\}$ around the north pole. Dissipation appears when Wick--rotating this axis by $\varepsilon< 0$ as done in (\ref{chois}) and changing variables as per (\ref{tiemponuevo}), in order to work in the coordinate chart $\left\{w\in\mathbb{C}\right\}$ around the south pole; this is how the $S$--picture of section \ref{babst} arises. The $H$--picture is purely conservative (because it satisfies the conservation law ${\rm d}H/{\rm d}t=0$), the $S$--picture is dissipative (because it satisfies the second law ${\rm Im}({\rm d}S/{\rm d}t)\geq 0$) . We realise that the $S$--picture involves dissipation/gravity, while the $H$--picture involves neither. This is analogous to the equivalence principle of gravitation, whereby the action of a gravitational field can be (locally) turned off by an appropriate change of coordinates. 

The foregoing arguments implement a relativity of the notion of {\it quantumness}\/ vs. {\it dissipation}\/ by means of ${\rm U}(1)$--transformations. However this ${\rm U}(1)$ symmetry of Wick rotations is broken the very moment one selects a specific value for $\varepsilon$. Hence the distinction between quantumness and dissipation (falsely) appears to be absolute, while in fact it is not. In particular, just as gravity can be (locally) gauged away, so can dissipation. Turn this argument around to conclude that {\it quantumness, or alternatively dissipation, can be gauged away}\/, although never the two of them simultaneously. Quantumness is gauged away in the limit $\varphi\to-\pi/2$, while dissipation is gauged away in the limit $\varphi\to 0$.\footnote{Since we have systematically dropped terms of order $\varepsilon^2$ and higher, some of our expressions may need amendments before taking the limit $\varphi\to-\pi/2$, but this does not invalidate our reasoning.} Moreover, our statement concerning the {\it relativity of dissipation}\/ is equivalent to our statement concerning the {\it relativity of quantumness}\/. A concept closely related to this latter notion was put forward in \cite{ROVELLI00}. Compare now the concept {\it relativity of quantumness}\/ with its transpose {\it quantum relativity}\/, or {\it quantum gravity}\/ as usually called: beyond the pun on words, these two concepts appear to be complementary, in Bohr's sense of the term ``complementarity".

\section{Discussion}\label{seis}

Our approach to quantum mechanics is an attempt to meet the requirement (demanded {\it e.g.}\/ in \cite{PENROSE3, SMOLIN1, SMOLIN2}, among others) that gravity be incorporated into the foundations of quantum theory. The absence of a link between quantum and gravitational effects in the standard formulation of quantum theory is a feature that has been claimed to lie at the heart of some of the conceptual difficulties facing the foundations of quantum mechanics. 

Specifically, in this paper we have presented a thermodynamical approach (following the classical theory of irreversible thermodynamics \cite{ONSAGERRECIP, ONSAGER, PRIGOGINE}) that provides a viable answer to this request, at least in a certain limit to be specified below. The incorporation of gravitational effects in a discussion of the principles of quantum mechanics is being addressed here through the appearance of dissipation as a gravitational effect. In this way the time--reversal symmetry of quantum mechanics is destroyed. Nonunitarity is implemented here by means of a Wick rotation; the latter is a consequence of gravitation.  In fact Wick rotations of the time axis are the quantum--mechanical counterpart to the equivalence principle of gravitation. Just as gravity can be (locally) gauged away, so can dissipation/quantumness.

For ease of reference, below we present Eqs. (\ref{tiemponuevo}), (\ref{iebolution}), (\ref{maniare}), (\ref{disipatiu}) and (\ref{incierto}) again in order to summarise the relevant expressions of the $S$--picture of quantum mechanics developed in this paper. We have
\begin{equation}
\frac{{\rm e}^{{\rm i}\varepsilon}}{t}=\frac{k_B}{\hbar}T,\qquad\tau=\ln \left(\frac{T}{T_0}\right), 
\label{noutemps}
\end{equation}
which relates inverse time and temperature through a Wick rotation by a small, dimensionless parameter $\varepsilon<0$. The latter encodes the strength of an external gravitational field; in the absence of gravitation we have $\varepsilon=0$. Applying the change of variables (\ref{noutemps}), the usual Schroedinger equation and the uncertainty principle become
\begin{equation}
k_B\frac{{\rm d}\psi}{{\rm d}\tau}=({\rm i}-\varepsilon)S\psi,\qquad S=\frac{H}{T},\qquad \Delta S\Delta \tau\gtrsim k_B,
\label{negmangiare}
\end{equation}
where the Hamiltonian operator $H$ is replaced with the entropy operator $S$. This entropic Schrodinger equation is solved by $\psi(\tau)={\cal S}(\tau)\psi(0)$, where the evolution operators ${\cal S}(\tau)$ in the dimensionless parameter $\tau$, defined as
\begin{equation}
{\cal S}(\tau):={\cal T}\exp\left(\frac{{\rm i}-\varepsilon}{k_B}\int_0^{\tau}S(\tilde\tau){\rm d}\tilde\tau\right),
\label{negiebolution}
\end{equation}
satisfy a 1--parameter semigroup of nonunitary operators (above, ${\cal T}$ denotes operator ordering along the parameter $\tilde\tau\geq 0$). Finally the expression
\begin{equation}
{\rm Im}\left(\frac{{\rm d}S}{{\rm d}t}\right)=-\varepsilon\frac{k_B}{\hbar}H
\label{negdisipatiu}
\end{equation}
relates the rate of entropy production to the Hamiltonian operator, while at the same time fixing the sign of $\varepsilon$ to be negative, in compliance with the second law of thermodynamics.

The previous equations hold in the limiting case of a weak gravitational field acting on a quantum particle described by the same equations. In view of the smallness of $\varepsilon$ in (\ref{negiebolution}), it is only for large values of $\tau$ that one can hope to measure the appearance of unitarity loss. It is important to realise that, by just switching back and forth between the energy picture (standard quantum mechanics) and the entropy picture (as summarised in Eqs. (\ref{noutemps}), (\ref{negmangiare}),  (\ref{negiebolution})  and (\ref{negdisipatiu})), either quantumness or dissipation can be gauged away, though never the two of them simultaneously. This fact we take as a reflection of the equivalence principle of relativity, whereby gravitational fields can be (locally) gauged away by means of coordinate changes.

The postulate (\ref{tiemponuevo}) (first presented long ago by de Broglie \cite{DEBROGLIE} without the Wick rotation ${\rm e}^{{\rm i}\varepsilon}$) leads to {\it considering time as emergent a property as temperature itself}\/. In this way unitarity violation can also be regarded as an emergent phenomenon.

\vskip.5cm
\noindent
{\bf Acknowledgements} J.M.I. would like to thank the organisers of the {\it Sixth International Workshop DICE 2012: Spacetime -- Matter -- Quantum Mechanics: from the Planck scale to emergent phenomena}\/ (Castiglioncello, Italy, September 2012),  for stimulating a congenial atmosphere of scientific exchange, and for the many interesting discussions that followed.\\
{\it Forse altro canter\`a con miglior plettro}---L. Ariosto.

\end{document}